\documentclass{article}
\usepackage{floatflt}
\usepackage{graphicx}
\usepackage{frascatiphys_SP}
\begin{document}
\title{ 
$K\to3\pi$ DECAYS IN CHIRAL PERTURBATION THEORY
}
\author{
Fredrik Borg \\
\em  Department of Theoretical Physics, Lund University,\\ 
\em S\"{o}lvegatan 14A,
SE - 223 62 Lund, Sweden
}
\maketitle
\baselineskip=11.6pt

\baselineskip=14pt
\section{\bf Introduction}
A kaon decaying into three pions is an example of a weak process. 
However, since the quarks 
are confined into mesons, the strong interaction also plays an
important part. The 
reaction is a low-energy one, meaning that it takes place in the 
non-perturbative region of QCD. In that region perturbative QCD doesn't 
give you any answers and other methods have to be used. The one we 
use is called Chiral Perturbation Theory. 
\section{\bf Chiral Perturbation Theory}
ChPT is an effective field theory describing the low-energy interactions of 
the kaons, the pions and the eta. It can be used for 
strong as well as non-leptonic weak interactions.
The  Chiral Lagrangian is based on the spontaneous breaking of 
chiral symmetry.

Chiral symmetry is the separate symmetry between the left- and right-handed 
quarks. In the theory only the u, d and s quarks are included which means 
that the symmetry group is $SU(3)_L \times SU(3)_R$. This symmetry is 
spontaneously broken by the vacuum condensate into $SU(3)$. Since the 
symmetry is only {\em approximate} (exact if $m_u= m_d= m_s=0$), this generates 
8 {\em light} (not massless) 
Goldstone particles, identified as the kaons, pions and eta. From the 
knowledge of interactions between Goldstone particles one then constructs 
the Chiral Lagrangian. 

The Chiral Lagrangian is organized in terms of importance. However, since 
it deals with low-energy processes, $\alpha_S$ can not be used for this 
purpose. Instead it is written as an expansion in $p$ and 
$m$, the momenta and the masses of the pseudoscalars $(K,\pi,\eta)$. 
Properly normalized 
these quantities 
are small and can be used as perturbation expansion parameters. Lowest order 
then means $p^2$ and $m^2$ and next-to-leading order $p^4$, $m^4$, $p^2m^2$ 
and so on. If one includes also isospin breaking, the unit charge, $e$, is 
also considered an expansion parameter.
\section{\bf Isospin Symmetry}
Calculations are often performed in the isospin limit, where the u and d 
quarks are treated as being identical. In practice this means setting 
$m_u= m_d$ and neglecting electromagnetism. 

In our first paper\cite{BDP} the calculation was made in the isospin 
limit. In the second paper\cite{BB} we took into account strong 
isospin breaking, ie.\ the quark mass 
difference $m_u-m_d$ as well as the local electromagnetic effects. 
Work is in progress to evaluate the other electromagnetic corrections as well. 
\section{\bf Results}
There are five different CP-conserving decays of the type $K\to3\pi$. The 
$K^-$ decays are not treated since they are counterparts
to the $K^+$ decays.
\begin{floatingfigure}[r]{40mm}
\begin{center}
\setlength{\unitlength}{1pt}
\resizebox{90 pt}{!}{
\rotatebox{0}{
\includegraphics{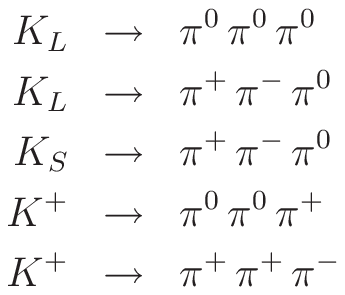}}
}
\end{center}
\end{floatingfigure}
A full isospin limit fit was made in \cite{BDP} taking into account all
data published before May 2002. One of the reasons for the further 
investigation 
of isospin breaking effects is to see whether isospin violation can solve 
the discrepancies in the quadratic slope parameters found there. A new full 
fit will
be done after all the electromagnetic contributions have been included
in the amplitudes (work in progress). 

\subsection{Results with and without strong isospin breaking}
Our main result up to now is the comparison between the amplitudes in the 
isospin limit
and including first order strong isospin breaking. 
In Fig.~\ref{fig:phase} we show the phase space boundaries for the five 
different decays and the three curves along which we compared
the squared amplitudes with and without first order strong isospin 
breaking

\begin{figure}
\begin{center}
\setlength{\unitlength}{1pt}
\resizebox{300 pt}{!}{
\rotatebox{270}{
\includegraphics{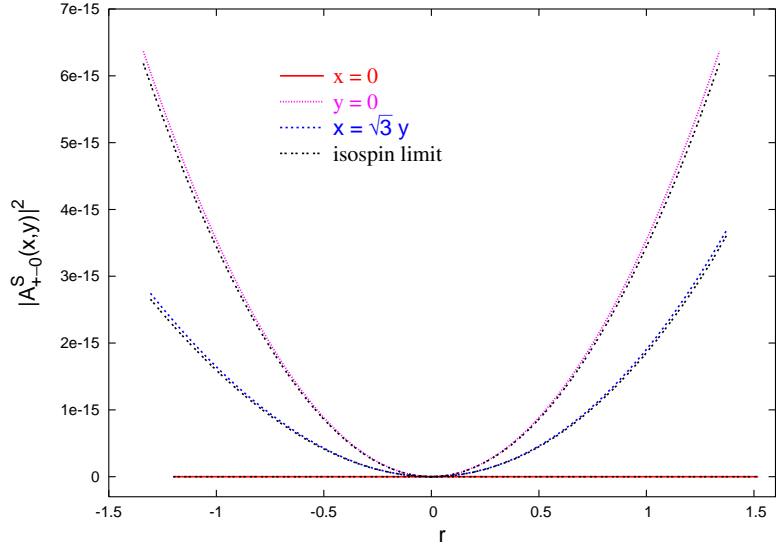}}}
\end{center}
\caption{\em $K_S\to\pi^+\pi^-\pi^0$ with and without 
strong isospin breaking.}
\label{fig:kspmo}
\end{figure}

In general the differences
are of the size to be expected from this type of isospin breaking.
For $K_L\to\pi^0\pi^0\pi^0$
the central value of the amplitude squared increases by about 3\% when
strong isospin breaking is included. 
The change in the quadratic slope is similar but the total variation over
the Dalitz plot is small so the total decay rate increases by about 3\% as 
well.
The squared amplitude  $K_L\to\pi^+\pi^-\pi^0$ increases by about
2.5\%. The decay rate and the changes in the Dalitz plot slopes are 
of similar size.
For the decay $K_S\to\pi^+\pi^-\pi^0$ the amplitude in the center of the Dalitz
plot vanishes because of CP-asymmetry. The amplitude and the slopes
increase by about 3\%, see Fig.~\ref{fig:kspmo}.
The decay $K^+\to\pi^0\pi^0\pi^+$ has the largest increase.
The squared amplitude
in the center changes by about 11\%. The linear slopes decrease somewhat
leading to an increase of about 8\% to the total decay rate when compared
with the isospin conserved case.
The decay $K^+\to\pi^+\pi^+\pi^-$ has a change of about 7.5\% upwards in
the center of the Dalitz plot and a similar change in the decay rate.
The slopes decrease somewhat. For more figures and detailed results, see 
\cite{BB}. 
\section{\bf Conclusions}
\begin{floatingfigure}[r]{67mm}
\begin{center}
\setlength{\unitlength}{1pt}
\resizebox{170 pt}{!}{
\rotatebox{270}{
\includegraphics{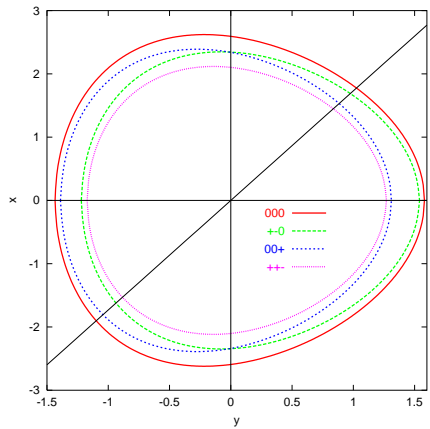}}
}
\end{center}
\caption{\em The phase space boundaries for the five different decays 
and the curves 
along which we will compare the amplitudes.\label{fig:phase}}
\end{floatingfigure}
We have calculated the $K\to3\pi$ amplitudes to
next-to-leading order in ChPT. 
A first calculation was done in \cite{BDP} in the isospin limit,
but we have now also included effects from $m_u\neq m_d$ and local
electromagnetic isospin breaking in \cite{BB}. 
This was done partly
because it is
interesting in general to see the possible importance of isospin breaking
in this process, but also to investigate whether isospin violation will
improve the fit to experimental data made in \cite{BDP}. 
We have tried to estimate the effects of the breaking by comparing
the squared amplitudes with and without isospin violation. The effect
seems to be at a few percent level, and probably not quite enough to solve the
dicrepancies. However, to really investigate this a new full fit has to be
done, including the explicit photon diagrams and the new data 
published
after \cite{BDP} as well. This is work in progress
and will be presented in future papers.

\section{Acknowledgements}
The program FORM 3.0 has been used extensively in these calculations. 
This work is supported in part by the Swedish Research Council
and European Union TMR
network, Contract No. HPRN-CT-2002-00311  (EURIDICE).

\end{document}